\documentclass[runningheads]{llncs}
\usepackage[T1]{fontenc}
\usepackage{graphicx}
\usepackage{amsmath}
\usepackage{color}
\usepackage[textwidth=6em,textsize=tiny]{todonotes}
\usepackage{color}

\usepackage{soul}
\usepackage[numbers,square]{natbib}

\begin{document}
\textbf{\textit{This manuscript has been published in Deep Generative Models. DGM4MICCAI 2024. Lecture Notes in Computer Science, vol 15224. Springer, Cham, 09 October 2024}} \\

\url{https://doi.org/10.1016/j.cpet.2021.07.001} \\

ISBN: 978-3-031-72744-3
\vspace{400pt}

\title{How To Segment in 3D Using 2D Models: Automated 3D Segmentation of Prostate Cancer Metastatic Lesions on PET Volumes Using Multi-Angle Maximum Intensity Projections and Diffusion Models}

\titlerunning{How To Segment in 3D Using 2D Models}

\author{Amirhosein Toosi \inst{1,5}         \orcidID{0000-0001-6432-9428}   
\and Sara Harsini        \inst{2}           \orcidID{0000-0001-6196-6982}   
\and François Bénard     \inst{3,5}         \orcidID{0000-0001-7995-3581}   
\and Carlos Uribe        \inst{1,4,5}       \orcidID{0000-0003-3127-7478}   
\and Arman Rahmim        \inst{1,5,6,7}     \orcidID{0000-0002-9980-2403}}  


%
\authorrunning{A. Toosi et al.}

\institute{ Department of Integrative Oncology, BC Cancer Research Institute, Vancouver, BC, Canada
\and BC Cancer Research Institute, Vancouver, BC, Canada
\and BC Cancer, Vancouver, BC, Canada
\and Functional Imaging, BC Cancer, Vancouver, BC, Canada
\and Department of Radiology, University of British Columbia, Vancouver, BC, Canada 
\and Department of Biomedical Engineering, University of British Columbia, Vancouver, BC, Canada
\and Department of Physics and Astronomy, University of British Columbia, Vancouver, BC, Canada }



%
\maketitle
%

\begin{abstract}
Prostate specific membrane antigen (PSMA) positron emission tomography/computed tomography (PET/CT) imaging provides a tremendously exciting frontier in visualization of prostate cancer (PCa) metastatic lesions. However, accurate segmentation of metastatic lesions is challenging due to low signal-to-noise ratios and variable sizes, shapes, and locations of the lesions. This study proposes a novel approach for automated segmentation of metastatic lesions in PSMA PET/CT 3D volumetric images using 2D denoising diffusion probabilistic models (DDPMs). Instead of 2D trans-axial slices or 3D volumes, the proposed approach segments the lesions on generated multi-angle maximum intensity projections (MA-MIPs) of the PSMA PET images, then obtains the final 3D segmentation masks from 3D ordered subset expectation maximization (OSEM)  reconstruction of 2D MA-MIPs segmentations. Our proposed method achieved superior performance compared to state-of-the-art 3D segmentation approaches in terms of accuracy and robustness in detecting and segmenting small metastatic PCa lesions. The proposed method has significant potential as a tool for quantitative analysis of metastatic burden in PCa patients. The code developed for this work is publicly available at: \url{https://github.com/Amirhosein2c/MIP-DDPM} .

\end{abstract}

\section{Introduction}
\label{sec:intro}

Prostate cancer (PCa) is the second most prevalent cancer and the fifth leading cause of cancer-related mortality among men \cite{soldatov2019patterns}. Despite progress in conclusive local treatments like radical prostatectomy (RP) and radiation therapy (RT), an estimated 20-50\% of patients will experience biochemical recurrence (BCR), marked by increasing levels of prostate-specific antigen (PSA). \cite{freedland2003time} 
The recurrence of prostate cancer may manifest as metastasis in the regional lymph nodes and bone structures. As the disease progresses, involvement of the liver and lungs, among other sites, may also occur \cite{bubendorf2000metastatic}. Depending on which site is involved with the disease, different type of treatment might be required. 

PSA level raise is considered as the primary biomarker for following up on prostate cancer treatment response and monitoring disease recurrence in prostate cancer patients \cite{duffy2020biomarkers}. However, it cannot localize the recurrence of the disease. Therefore, the precise identification of recurrence locations becomes significantly important for therapeutic decision-making processes. 
As a result, employing a diagnostic imaging modality that possesses both high sensitivity and specificity is crucial for differentiating between local relapse, oligometastatic disease, and extensive disease, hence to enable individualized treatment plans for patients. 
Recent advancements in Positron Emission Tomography (PET) imaging has led to improved detection and quantification of many types of primary and metastatic lesions. Design of recent PET radiopharmaceuticals that are able to target the prostate-specific membrane antigen (PSMA), such as [\textsuperscript{18}F]DCFPyL, with much higher sensitivity and specificity compared to conventional imaging modalities has opened a new era in the diagnosis, treatment decision-making, and patient management in prostate cancer \cite{rousseau2019prospective}. 

Deep learning algorithms, have shown great potential in computer-aided diagnosis \cite{ma2022clinical}. Yet, challenges arise from the nature of the imaging modality that AI-based image recognition algorithms must cope with, including low contrast, large intra- and inter-patient heterogeneity, and blurring and noise in the images. The unique specifications of biochemical recurrent prostate cancer metastatic lesions, including tumors with very small sizes, low-to-moderate radiopharmaceutical uptake, especially compared to the high biological uptake of the bladder and kidneys, make them hard targets to detect \cite{fendler201768}. Added to that, local tumor recurrence adjacent to the urinary bladder, or in the abdomen area with high background noise and high uptake regions such as ureters, further complicate the detection of PCa lesions even for physicians, making the task of manual segmentation time and labor-intensive \cite{haupt202068}. As such, localizing the lesions in the image could help physicians save time and increase the accuracy of the task.

 There are limited number of works that tried to tackle the problem of PCa tumor/metastatic lesion detection and segmentation using the power of AI. Prior works mainly focused on the local primary (intra-prostatic) tumor segmentation \cite{kostyszyn2021intraprostatic} which is a relatively less challenging task, given the locality of the disease occurrence. Only in \cite{jafari2023convolutional} and \cite{xu2023automatic} authors evaluated the performance of CNN-based segmentation models for PCa lesions segmentation, however only the dataset used in \cite{xu2023automatic} is for PCa recurrence patients.

In this work, we introduce an innovative approach for automated detection and segmentation of biochemically recurrent PCa metastatic lesions on PSMA-PET images using a 2D Diffusion-based segmentation model. This approach includes novel use of the ordered-subset expectation-maximization (OSEM) algorithm applied to 2D segmentations of multi-angle maximum intensity projections (MA-MIPs) to generate 3D segmentation of metastatic lesions in PET images, while taking advantage of the computational efficiency and performance benefits of training a 2D diffusion-based segmentation model on MA-MIPs. We show that our method outperforms its state-of-the-art 3D rivals in terms of various segmentation metrics on the target dataset. The code developed for this work is publicly available at: \url{https://github.com/Amirhosein2c/MIP-DDPM} .

\section{Methods and Materials}
\subsection{Dataset}
\label{sec:dataset}
This is a post‑hoc sub-group analysis of a prospective clinical trial. Inclusion criteria were: (1) histologically proven prostate cancer with biochemical recurrence after initial curative therapy with radical prostatectomy, with a PSA > 0.4 ng/mL and an additional measurement showing increase; (2) histologically proven prostate cancer with biochemical recurrence after initial curative therapy with RT, with a PSA level > 2 ng/mL above the nadir after therapy. \cite{harsini2023outcome}. Overall, $510$ whole-body [\textsuperscript{18}F]DCFPyL PSMA-PET/CT images were chosen. Each trans-axial PET image has a matrix size of $192 \times 192$ pixels, with each pixel covering $3.64 mm^2$ in physical space. All active lesions were manually delineated by an expert nuclear medicine physician. On average each image had $1.92 \pm 1.21$ PCa lesions with an average active volume of $4.03\pm7.02 ml$ and long axis diameter of $12.96\pm10.11 mm$ (on CT). 
The average maximum standard uptake value (SUVmax) and SUVmean of all the lesions were $9.64\pm10.04$ and $4.4\pm3.55$, respectively.

\subsection{Data Preprocessing}
\label{sec:Preprocessing_Augmentation}
PSMA-PET activity concentration values (Bq/ml) of all PSMA PET voxels were converted to Standard Uptake Value (SUV). To decrease the contrast between high uptake normal organs and the small lesions, SUV values were clipped to a range of 0 to 25. 
CT images had an original voxel size of $(0.98\times0.98\times3.27) mm^3$, and the PET images had a voxel size of $(3.64\times3.64\times3.27) mm^3$. All PET/CT images were resampled to a voxel size of $(2.0\times2.0\times2.0) mm^3$ using a third-order spline method for both CT and PET images and further cropped to have matrix size of $250\times250$. 72 axial rotations of the PSMA PET volumes were computed in every $5^{\circ}$ degrees of axial rotation, and the maximum intensity projections (MIPs) of all 72 volumes (the original volume and all 71 rotated ones) were computed, in order to cover one complete $360^{\circ}$ axial rotation of the volume.
Since preserving the information of soft tissues in the CT equivalent of MIP projections are not feasible, in order to provide more context information to the DDPM network, 4 different MIP projections per each axial rotation were computed: First, the normal MIP taking by computing the maximum value of each ray in the rotated 3D volume (PET-MIP); second, projecting the maximum intensity after further clipping the PSMA-PET voxel to the range of 0 - 10 (PET10-MIP); third, further clipping the voxels to the range of 0 - 5 and capturing MIP (PET5-MIP); and finally, the depth location of the voxels with maximum intensity along each ray (DEPTH-MIP). To the best of our knowledge, this is the first time that such derived modality is being used for segmentation in the related literature.

\subsection{Segmentation Network Architecture}
\label{sec:DDPM_Architecture}
The model used in this work for automated segmentation of the metastatic lesions on MA-MIPs is a denoising diffusion probabilistic model (DDPM) initially proposed in \cite{nichol2021improved} and modified in \cite{wolleb2022diffusion} for brain tumor segmentation on 2D trans-axial MRI slices. The general idea behind diffusion models comprised of two chains of incrementally noising and denoising, known as forward $(q)$ and reverse $(p)$ processes, respectively. The forward process $p$ starts with adding a small amount of Gaussian noise to the input image $x$ over $T$ time steps, resulting in a series of noisy images $x_0,x_1,\ldots,x_T$. Then, during the reverse process $p$, the model which is a U-Net architecture based network, learns to predict the slightly less noisy image $x_{t-1}$ from $x_t$ for each step $t \in \{1, \ldots, T\}$ . Throughout the training of the diffusion model, the ground truth image $x_{t-1}$ in each time step $t$ is known, and hence the model can be trained using $L_2$ loss. 

During test time, the sampling process $p$ starts from random Gaussian noise $x_T \sim \mathcal{N}(0, \mathbf{I})$, and iteratively denoises it using the trained U-Net model to generate a fake image $x_0$. 

Writing the forward process $q$ as $q(x_t|x_0) :=\mathcal{N}(x_t;\sqrt{\bar{\alpha}_t}x_0,(1-\bar{\alpha}_t)\mathbf{I})$ where $\alpha_t:=1-\beta_t$ and $\beta_t$ is the variance of the forward process $q$ at the time step $t$, and $\bar{\alpha}_t:=\prod_{s=1}^{t}\alpha_s$, then $x_t$ can be directly expressed based on $x_0$ as:
\begin{equation} \label{eq:1}
   x_t = \sqrt{\bar{\alpha}_t}x_0 + \sqrt{1-\bar{\alpha}_t}\epsilon, \ \ \ \text{where} \ \ \epsilon \sim \mathcal{N}(0, \mathbf{I}) 
\end{equation}
as shown in \cite{ho2020denoising}.
For the reverse (denoising) process, given the parameters of the trained U-Net model ($\theta$), the learned reverse process $p_\theta$ can be written as $p_\theta(x_{t-1}|x_t):=\mathcal{N}(x_{t-1};\mu_\theta(x_t, t),\Sigma_\theta(x_t,t))$. Here $x_{t-1}$ can be predicted using the following formula as given in \cite{ho2020denoising}:

\begin{equation} \label{eq:2}
    x_{t-1} = 1\frac{1}{\sqrt{\alpha_t}} \left(x_t - \frac{1-\alpha t}{\sqrt{1-\bar{\alpha}_t}} \epsilon_\theta (x_t, t)\right) + \sigma_t \mathbf{z}, \ \ \ \text{where} \ \ \mathbf{z} \sim \mathcal{N}(0,\mathbf{I})
\end{equation}
$\sigma_t$ is the variance scheme of the reverse process \cite{nichol2021improved} and $\mathbf{z}$ is the random component of the sampling process. The U-Net denoted as $\epsilon_{\theta}$ at each time step $t$ takes $x_t$ (as defined in equation\ref{eq:1}) as input, and learns the noise scheme $\epsilon_{\theta}(x_t, t)$. At the time step $t$ during sampling, the predicted $\epsilon_{\theta}$ is subtracted from $x_t$ according to Equation \ref{eq:2} to construct $x_{t-1}$ which is a slightly less noisy version of the input $x_t$.

\begin{figure}
\includegraphics[clip, width=\textwidth,trim=4.0cm 60.0cm 0.0cm 10.0cm]{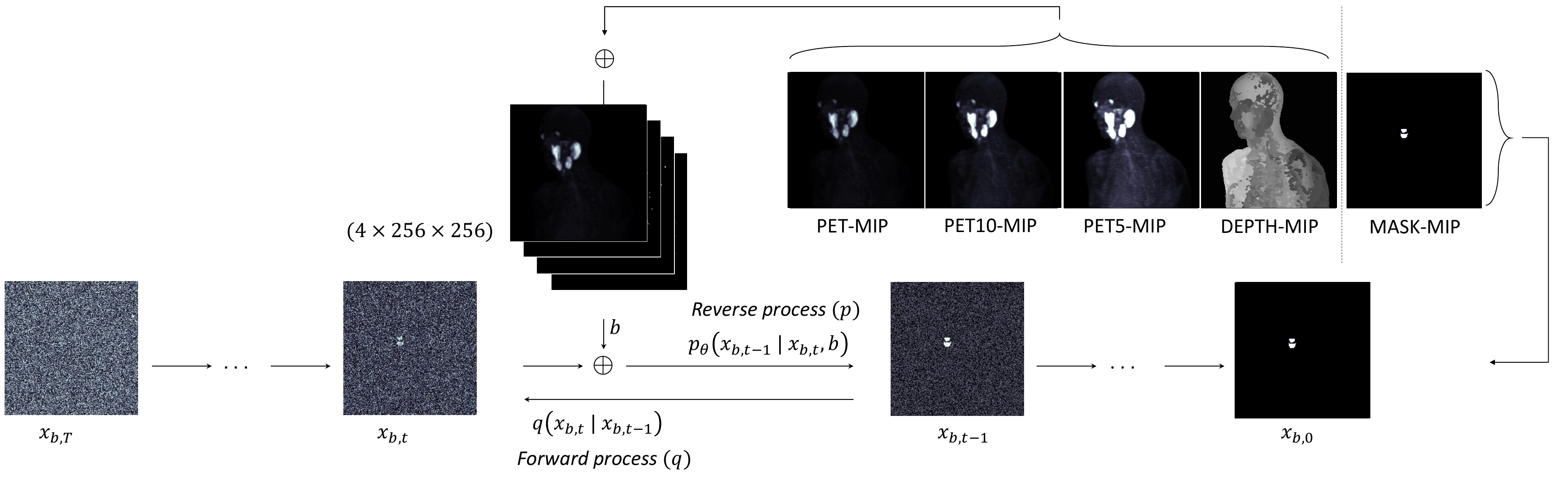}
\caption{ Visual explanation of how the forward and reverse processes of DDPM model works, along with the input ground truth segmentation mask during the forward process and the prior information of the anatomical/functional context during the reverse process} \label{fig:overview}
\end{figure}

In its classic form, a diffusion model trained on a dataset, during the inference time, takes a random Gaussian noise $x_T$ as input, and generates a synthetic image that fits in the distribution of the training dataset. In case of semantic segmentation, starting from a random noise, the trained diffusion model will generate a mask according to the distribution of the segmentation masks it was trained on, but not necessarily the segmentation mask of the test sample it has been given. As such, in \cite{wolleb2022diffusion} the authors modified the forward and reverse processes of a classic diffusion model by providing the brain MR axial slices as prior information and binding them to the anatomical information. As denoted in \cite{wolleb2022diffusion}, the trans-axial image $b$ corresponding to the ground truth segmentation mask $x_b$ are concatenated together, $X := b\oplus x_b$. The incremental noise during the forward process $q$ is only added to the ground truth segmentation mask $x_b$, defined as $x_{b,t}$. As a result,  equation \ref{eq:1} is modified as follows:
\begin{equation}\label{eq:3}
   x_{b,t} = \sqrt{\bar{\alpha}_t}x_b + \sqrt{1-\bar{\alpha}_t}\epsilon, \ \ \ \text{where} \ \ \epsilon \sim \mathcal{N}(0, \mathbf{I}) 
\end{equation}
Accordingly,  equation \ref{eq:2} of the reverse process $p$ is rewritten as follows:
\begin{equation} \label{eq:4}
    x_{b,t-1} = 1\frac{1}{\sqrt{\alpha_t}} \left(x_{b,t} - \frac{1-\alpha t}{\sqrt{1-\bar{\alpha}_t}} \epsilon_\theta (X_t, t)\right) + \sigma_t \mathbf{z}, \ \ \ \text{where} \ \ \mathbf{z} \sim \mathcal{N}(0,\mathbf{I})
\end{equation}
Here the input image prior $b$ is of size $(c, h, w)$, where each channel has the size of $h\times w$. The corresponding ground truth segmentation mask $x_b$ is of size $(1, h, w)$. Consequently, X has dimension $(c+1, h, w)$. Figure \ref{fig:overview} visually summarizes the forward and reverse processes during training of the DDPM model for the task of segmentation.

The stochastic property of the DDPM enables us to generate more than one segmentation mask prediction per each input image, thus, ensembling the multiple predictions of the same input image can potentially improve the segmentation performance. Therefore, during sampling we ensemble 10 segmentation mask variations and take the mean image as the predicted segmentation for the input MA-MIP.

\subsection{3D Reconstruction of 2D Masks Using OSEM Algorithm}
\label{sec:reconstruction}

The Ordered Subsets Expectation Maximization (OSEM) algorithm is an iterative reconstruction technique widely used in medical imaging modalities such as Positron Emission Tomography (PET) and Single-Photon Emission Computed Tomography (SPECT) \cite{hudson1994accelerated}. It reconstructs 3D volumetric images from a series of 2D tomographic projection data acquired at different angles around the subject. OSEM is an accelerated variant of the Expectation Maximization (EM) algorithm \cite{moon1996expectation}, which aims to estimate the unknown 3D radio-tracer distribution within the subject by maximizing the likelihood of observing the measured 2D projection data. The OSEM algorithm introduces an ordered subsets approach to accelerate the convergence rate of the EM algorithm. 

A novelty of our work is that OSEM algorithm, unlike conventionally applied to acquired data to generate images, is applied to segmentations of MA-MIPs as generated from the images. By utilizing the OSEM algorithm, the proposed method in this study efficiently reconstructs the 3D segmentation volume from the predicted 2D segmentation masks obtained from the MA-MIPs. The back-projection step of the OSEM algorithm is used to map the 2D segmentation masks onto the 3D volume, iteratively refining the estimate until convergence.

\section{Experiment Details}
\label{sec:experiment_detail}

We evaluated our proposed MIP-based segmentation method on the PSMA-PET image dataset described in section \ref{sec:dataset}. 
Per each patient, 72 axial rotated volumes of the PSMA-PET image were generated. In order to avoid padding the images as much as possible and providing more meaningful information to the network, all the whole-body PSMA-PET volumes were divided into two section, upper body and lower body, resulting in 144 volumes per each patient. For each volume, 4 different MIP images were generated, as described in section \ref{sec:Preprocessing_Augmentation}. These four different MIPs were stacked to prepare the input data for training the DDPM model, resulting in the size of $(4, 250, 250)$. 
These steps resulted in 66240 images from 460 PSMA-PET volumes of the same number of patients for training/validation of the DDPM model and 7200 images for testing from 50 patients. 

Backbone of the DDPM model used in this work is a U-Net architecture network described in \cite{wolleb2022diffusion}, and \cite{nichol2021improved} with input size of five-channel $256\times256$. It uses six feature map with resolutions from $128\times128$ to $4\times4$, two residual blocks, and one self-attention head with $16\times16$ resolution. Similar to \cite{wolleb2022diffusion} we chose a 10000-steps linear noise schedule for training/sampling. The model is trained for 72 hours (150,000 iterations) on an NVIDIA V100 GPU 16 GB, with a learning rate of $10^{-4}$ using Adam optimizer, and a batch size of 1. 

In order to benefit from the ensemble of multiple predicted masks using the stochastic property of the trained DDPM model, for each image in the test set, 10 variations of segmentation masks were predicted. The computed mean image of the 10 masks were taken as the final segmentation per each projection, after thresholding each of the 10 predicted mask at $0.5$.

Next, the 72 segmentation masks per each volume in the test dataset were reconstructed to 3D using the OSEM algorithm, 
utilizing the PyTomography toolbox \cite{polson2023pytomography}. For each 3D reconstructed mask, OSEM ran for 40 iterations and 20 subsets, which took around 3 minutes per each volume. Finally per each patient in the test set, the two volumes of upper and lower body were stitched together to make the whole-body segmentation mask. 

In order to evaluate the performance of our proposed method, we trained eight 3D-based segmentation methods among the state-of-the-art biomedical image segmentation networks in the literature. Same dataset (PET and CT volumes, before MA-MIP generation) were used for training and evaluating these models. The input size of all methods were two-channel cropped volumes of size $128\times128\times128$. Batch sizes, number of crops, and other hyper-parameters were modified in a such way to let the 3D networks fit on four NVIDIA V100 16GB gpus for training. Sliding window inferencing with size of $192\times192\times192$ were used in test time for all models except the ones with explicit requirement of having the same size of training sample crops, $128\times128\times128$. All models were trained for 500 epochs and best model based on the validation set loss were picked for inferencing. Implementations were done using MONAI toolbox, python 3.10, and PyTorch 2.1, on Ubuntu 18.04 LTS. 
Results are reported briefly in table \ref{tab:results_main} of the next section.

\section{Results and Discussion}
\label{sec:result_discussion}

In this section we report the result of the proposed MA-MIP based segmentation method on our test set. As baseline  comparison, we also show results for 8 state-of-the-art (SOTA) 3D segmentation methods  in the literature.

\begin{table*}[h]
\caption{Performance evaluation of our method compared to the SOTA methods}
    \centering
    \begin{tabular}{p{3.2cm}p{1.8cm}p{1.8cm}p{1.8cm}p{2.5cm}} 

         \textbf{Model}   &\textbf{Dice}$\uparrow$    &\textbf{HD95}$\downarrow$&\textbf{Jaccard}$\uparrow$& \textbf{\%Vol. Error}$\downarrow$   \\ \hline
        
         U-Net                   &  ${0.461}$         &  ${45.32}$         &  ${0.345}$            &  ${45.6\%}$             \\ 
         U-Net++                 &  ${0.429}$         &  ${126.44}$        &  ${0.308}$            &  ${39.4\%}^{\star}$     \\ 
         Flexible U-Net (b1)     &  ${0.451}$         &  ${26.06}^{\star}$ &  ${0.343}$            &  ${48.5\%}$             \\ 
         Attention U-Net         &  ${0.467}$         &  ${135.77}$        &  ${0.337}$            &  ${48.4\%}$             \\ 
         SegResNet               &  ${0.470}^{\star}$ &  ${64.74}$         &  ${0.359}^{\star}$    &  ${42.3\%}$             \\ 
         UNETR                   &  ${0.407}$         &  ${80.01}$         &  ${0.299}$            &  ${45.1\%}$             \\ 
         Swin UNETR              &  ${0.438}$         &  ${56.60}$         &  ${0.328}$            &  ${43.6\%}$             \\ 
         V-Net                   &  ${0.426}$         &  ${57.80}$         &  ${0.325}$            &  ${51.7\%}$             \\ \hline
        \textbf{MIP-DDPM} (Ours) &  $\mathbf{0.532}$  &  $\mathbf{19.60}$  &  $\mathbf{0.433}$     &  $\mathbf{32.9\%}$      \\

    \end{tabular}
    \label{tab:results_main}
\end{table*}


\begin{figure}
\includegraphics[clip, width=\textwidth,trim=0.0cm 50.0cm 0.0cm 0.0cm]{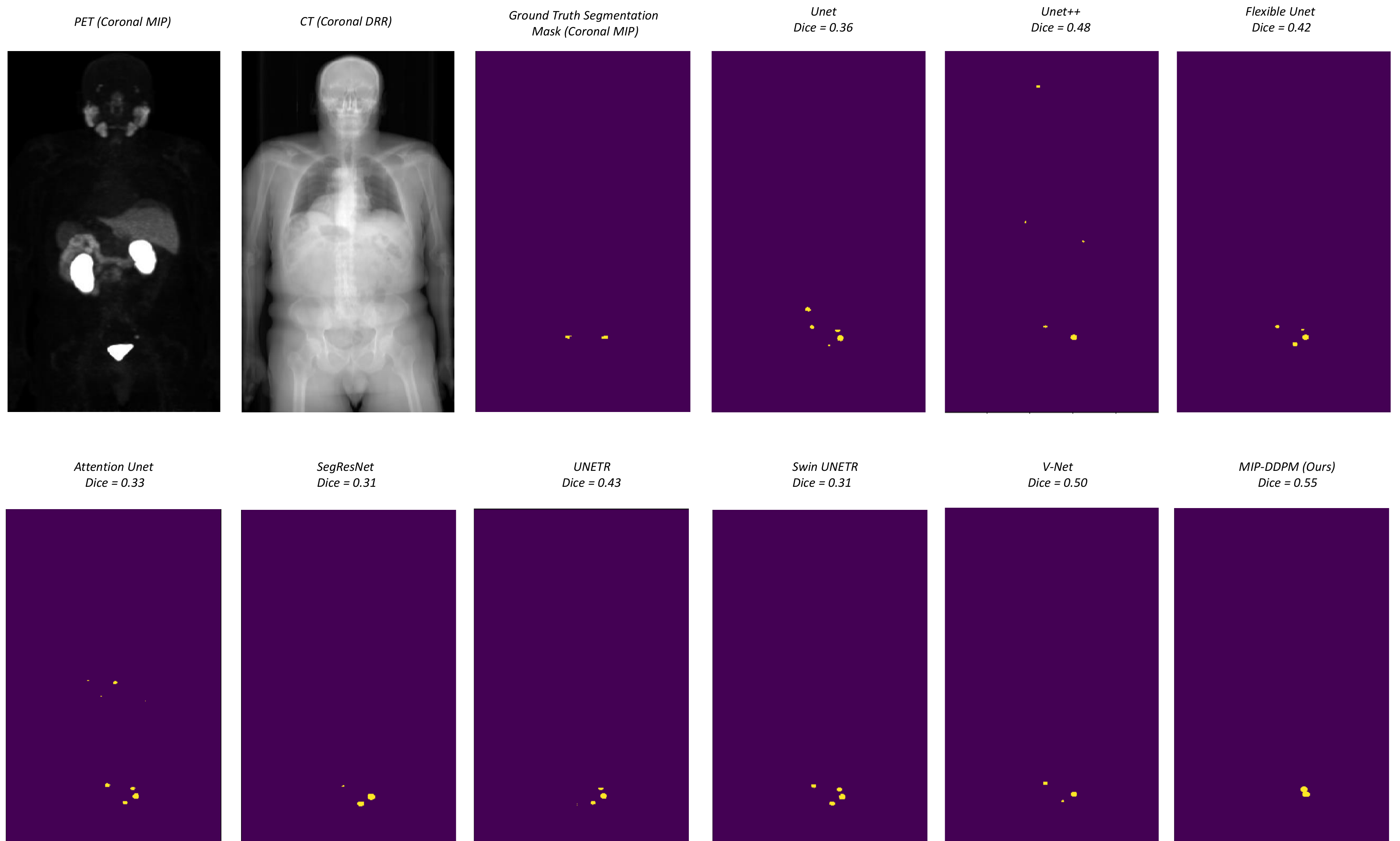}
\caption{ Visual comparison of our method against SOTA on a sample case. } \label{fig:result}
\end{figure}

Figure \ref{fig:result} shows predicted segmentation masks using the proposed and 8 SOTA techniques for a random test sample, also confirming visual improvements. Table \ref{tab:results_main} summarizes the overall results of our proposed method in terms of Dice score, 95 percentile Hausdorff Distance, Jaccard index, and Volume Error Percentage. It is seen that our proposed MIP-DDPM model outperforms all SOTA methods. 
PSMA PET images may enable earlier detection of metastases, which can improve management decisions, especially in case of biochemical recurrent disease. Hence, the primary clinical objective revolves around detecting both local recurrence and distant metastases. In many cases, as demonstrated in this study, the total tumor volume is small,  further challenging this task. Evaluating 17 SOTA object detection methods on the same dataset, the best performing model achieved 0.75 recall \cite{toosi2023advanced, toosi2023state}. 
In fact, on the dataset used in this study, a best performing prior automated segmentation method had achieved a mean dice score of only 0.38 using a SOTA self-supervised pre-training method \cite{xu2023automatic}. A very recent work on this dataset using SOTA 3D segmentation methods achieved a mean dice score of $\sim$ 0.47 using custom volume-preserving loss functions \cite{dzikunu20243d}. These  depict the complexity of the task, and that our proposed framework in this work  outperforms SOTA techniques previously reported. 

While our approach demonstrates promising results, it is important to consider some limitations and areas for future exploration. First, although we focused on segmentation metrics in this study, we recognize the potential value of incorporating detectability metrics, especially when dealing with small objects. This could provide additional insights into our method's performance.
The computational time of our method is an aspect that warrants further investigation. Accuracy is paramount in clinical settings, though, understanding and optimizing the time required for all the involved steps could enhance the practical applicability.

\section{Conclusion}

Our work introduces a novel method for segmenting small metastatic lesions on MA-MIPs using 2D DDPMs, demonstrating superior performance compared to state-of-the-art 3D segmentation methods.
The proposed method represents a significant step forward in addressing the challenging clinical problem of identifying small metastatic lesions.
Future directions for this research will include conducting a comprehensive ablation study to demonstrate the impact of each component in the proposed approach, exploring comparisons with other 2D and 3D diffusion-based methods, and validating on larger, multi-center datasets. 
By continuing to refine and expand this work, we aim to further improve the accuracy and efficiency of segmenting small objects in medical images, ultimately contributing to better patient care.

\begin{credits}
\subsubsection{\ackname} This work was supported by CIHR Project Grant PJT-162216, as well as computational resources provided by Microsoft AI for Health. 

\subsubsection{\discintname}
The authors have no competing interests to declare that are relevant to the content of this article.
\end{credits}

\bibliographystyle{splncs04.bst}
\bibliography{Paper-0043.bib}

\begin{thebibliography}{10}
\providecommand{\url}[1]{\texttt{#1}}
\providecommand{\urlprefix}{URL }
\providecommand{\doi}[1]{https://doi.org/#1}

\bibitem{bubendorf2000metastatic}
Bubendorf, L., Sch{\"o}pfer, A., Wagner, U., Sauter, G., Moch, H., Willi, N., Gasser, T.C., Mihatsch, M.J.: Metastatic patterns of prostate cancer: an autopsy study of 1,589 patients. Human pathology  \textbf{31}(5),  578--583 (2000)

\bibitem{duffy2020biomarkers}
Duffy, M.J.: Biomarkers for prostate cancer: prostate-specific antigen and beyond. Clinical Chemistry and Laboratory Medicine (CCLM)  \textbf{58}(3),  326--339 (2020)

\bibitem{dzikunu20243d}
Dzikunu, O., Ahamed, S., Toosi, A., Harsini, S., Benard, F., Rahmim, A., Uribe, C.: A 3d unet for automated metastatic lesions detection and segmentation from psma-pet images of patients with biochemical recurrence prostate cancer (2024)

\bibitem{fendler201768}
Fendler, W.P., Eiber, M., Beheshti, M., Bomanji, J., Ceci, F., Cho, S., Giesel, F., Haberkorn, U., Hope, T.A., Kopka, K., et~al.: 68 ga-psma pet/ct: Joint eanm and snmmi procedure guideline for prostate cancer imaging: version 1.0. European journal of nuclear medicine and molecular imaging  \textbf{44},  1014--1024 (2017)

\bibitem{freedland2003time}
Freedland, S.J., Presti~Jr, J.C., Amling, C.L., Kane, C.J., Aronson, W.J., Dorey, F., Terris, M.K., Group, S.D.S., et~al.: Time trends in biochemical recurrence after radical prostatectomy: results of the search database. Urology  \textbf{61}(4),  736--741 (2003)

\bibitem{harsini2023outcome}
Harsini, S., Wilson, D., Saprunoff, H., Allan, H., Gleave, M., Goldenberg, L., Chi, K.N., Kim-Sing, C., Tyldesley, S., B{\'e}nard, F.: Outcome of patients with biochemical recurrence of prostate cancer after psma pet/ct-directed radiotherapy or surgery without systemic therapy. Cancer Imaging  \textbf{23}(1), ~27 (2023)

\bibitem{haupt202068}
Haupt, F., Dijkstra, L., Alberts, I., Sachpekidis, C., Fech, V., Boxler, S., Gross, T., Holland-Letz, T., Zacho, H.D., Haberkorn, U., et~al.: 68 ga-psma-11 pet/ct in patients with recurrent prostate cancer—a modified protocol compared with the common protocol. European journal of nuclear medicine and molecular imaging  \textbf{47},  624--631 (2020)

\bibitem{ho2020denoising}
Ho, J., Jain, A., Abbeel, P.: Denoising diffusion probabilistic models. Advances in neural information processing systems  \textbf{33},  6840--6851 (2020)

\bibitem{hudson1994accelerated}
Hudson, H.M., Larkin, R.S.: Accelerated image reconstruction using ordered subsets of projection data. IEEE transactions on medical imaging  \textbf{13}(4),  601--609 (1994)

\bibitem{jafari2023convolutional}
Jafari, E., Zarei, A., Dadgar, H., Keshavarz, A., Manafi-Farid, R., Rostami, H., Assadi, M.: A convolutional neural network--based system for fully automatic segmentation of whole-body [68ga] ga-psma pet images in prostate cancer. European Journal of Nuclear Medicine and Molecular Imaging pp. 1--12 (2023)

\bibitem{kostyszyn2021intraprostatic}
Kostyszyn, D., Fechter, T., Bartl, N., Grosu, A.L., Gratzke, C., Sigle, A., Mix, M., Ruf, J., Fassbender, T.F., Kiefer, S., et~al.: Intraprostatic tumor segmentation on psma pet images in patients with primary prostate cancer with a convolutional neural network. Journal of Nuclear Medicine  \textbf{62}(6),  823--828 (2021)

\bibitem{ma2022clinical}
Ma, K., Harmon, S.A., Klyuzhin, I.S., Rahmim, A., Turkbey, B.: Clinical application of artificial intelligence in positron emission tomography: Imaging of prostate cancer. PET clinics  \textbf{17}(1),  137--143 (2022)

\bibitem{moon1996expectation}
Moon, T.K.: The expectation-maximization algorithm. IEEE Signal processing magazine  \textbf{13}(6),  47--60 (1996)

\bibitem{nichol2021improved}
Nichol, A.Q., Dhariwal, P.: Improved denoising diffusion probabilistic models. In: International Conference on Machine Learning. pp. 8162--8171. PMLR (2021)

\bibitem{polson2023pytomography}
Polson, L., Fedrigo, R., Li, C., Sabouri, M., Dzikunu, O., Ahamed, S., Rahmim, A., Uribe, C.: Pytomography: A python library for quantitative medical image reconstruction. arXiv preprint arXiv:2309.01977  (2023)

\bibitem{rousseau2019prospective}
Rousseau, E., Wilson, D., Lacroix-Poisson, F., Krauze, A., Chi, K., Gleave, M., McKenzie, M., Tyldesley, S., Goldenberg, S.L., B{\'e}nard, F.: A prospective study on 18f-dcfpyl psma pet/ct imaging in biochemical recurrence of prostate cancer. Journal of Nuclear Medicine  \textbf{60}(11),  1587--1593 (2019)

\bibitem{soldatov2019patterns}
Soldatov, A., von Klot, C.A., Walacides, D., Derlin, T., Bengel, F.M., Ross, T.L., Wester, H.J., Derlin, K., Kuczyk, M.A., Christiansen, H., et~al.: Patterns of progression after 68ga-psma-ligand pet/ct-guided radiation therapy for recurrent prostate cancer. International Journal of Radiation Oncology* Biology* Physics  \textbf{103}(1),  95--104 (2019)

\bibitem{toosi2023state}
Toosi, A., Harsini, S., Ahamed, S., Yousefirizi, F., B{\'e}nard, F., Uribe, C., Rahmim, A.: State-of-the-art object detection algorithms for small lesion detection in psma pet: use of rotational maximum intensity projection (mip) images. In: Medical Imaging 2023: Image Processing. vol. 12464, pp. 771--778. SPIE (2023)

\bibitem{toosi2023advanced}
Toosi, A., Harsini, S., Benard, F., Uribe, C., Rahmim, A.: Advanced deep learning-based lesion detection on rotational 2d maximum intensity projection (mip) images coupled with reverse mapping to the 3d pet domain (2023)

\bibitem{wolleb2022diffusion}
Wolleb, J., Sandk{\"u}hler, R., Bieder, F., Valmaggia, P., Cattin, P.C.: Diffusion models for implicit image segmentation ensembles. In: International Conference on Medical Imaging with Deep Learning. pp. 1336--1348. PMLR (2022)

\bibitem{xu2023automatic}
Xu, Y., Klyuzhin, I., Harsini, S., Ortiz, A., Zhang, S., B{\'e}nard, F., Dodhia, R., Uribe, C.F., Rahmim, A., Ferres, J.L.: Automatic segmentation of prostate cancer metastases in psma pet/ct images using deep neural networks with weighted batch-wise dice loss. Computers in Biology and Medicine  \textbf{158},  106882 (2023)

\end{thebibliography}

\end{document}